# The ERE of the "Red Rectangle" revisited. *

Hans Van Winckel[1]**, Martin Cohen[2], and Theodore R. Gull[3]

[1] Instituut voor Sterrenkunde, K.U.Leuven, Celestijnenlaan 200B, B-3000 Leuven, Belgium
[2] Radio Astronomy Laboratory, 601 Campbell Hall, University of California, Berkeley, CA 94720, USA
[3] Goddard Space Flight Center, Code 681, Greenbelt, MD 20771, USA

**Abstract.** We present in this paper high signal-to-noise long-slit optical spectra of the Extended Red Emission (ERE) in the "Red Rectangle" (RR) nebula. These spectra, obtained at different positions in the nebula, reveal an extremely complex emission pattern on top of the broad ERE continuum. It is well known that three features converge at large distance from the central object, in wavelength and profile to the diffuse interstellar bands (DIBs) at $\lambda$5797, $\lambda$5849.8, and $\lambda$6614 Å (e.g. Sarre et al., 1995). In this paper we give a detailed inventory of all spectral subfeatures observed in the 5550–6850 Å spectral range. Thanks to our high S/N spectra, we propose 5 new features in the RR that can be associated with DIBs. For the 5550–6200 Å spectral range our slit position was on top of the NE spike of the X shaped nebula. A detailed description of the spatial profile-changes is given of the strongest features revealing that even far out in the nebula at 24″ from the central star, there remains a small shift in wavelength of 1 respectively 2 Å between the ERE subfeatures and the DIB wavelengths of $\lambda$5797.11 and $\lambda$5849.78 Å.

## 1. Introduction

Since its identification by Cohen et al. (1975) the Red Rectangle nebula (RR) has remained a very popular target thanks to the many remarkable phenomena it displays across a very wide wavelength range. The RR has often been used as an archetypical example of a C-rich post-AGB object but it has recently become clear that many of the remarkable phenomena observed in the RR are closely related to the special binary nature of the central object, HD 44179, with its 318-day period and surprisingly high eccentricity (Van Winckel et al., 1995), and to the presence of a circumbinary disc (e.g. Waelkens et al., 1996). This disc is resolved in ground-based high-spatial-resolution imaging at optical and near-IR wavelengths (e.g. Roddier et al., 1995; Osterbart et al., 1997) as well as in HST snapshots (Bond et al., 1997). The central object is not seen directly at UV and optical wavelengths but only in the scattered light that escapes from the poles of the thick disc into the line-of-sight (Waelkens et al., 1996). The very high $L_{IR}/L_{UV-Opt.}$ ratio of about 33 (Leinert & Haas, 1989) despite the inferred low extinction towards the bright central object can also be explained by this structure.

We give a short synopsis of the rich observational characteristics of both the central star and the nebula.

HD 44179 ($m_v$=9) is severely depleted in refractory elements while volatile elements have nearly solar abundances (Waelkens et al., 1996). The most likely situation to mimic this ISM-like depletion of refractories is when circumstellar material is trapped in a long-lived circumbinary reservoir (Waters et al., 1992).

The extremely narrow, weak, microwave CO emission observed in the RR (Jura et al. 1995) and the inferred presence of very large grains (Jura et al., 1997), are distinct characteristics also best explained by long-term processing of dust in a circumbinary disc. Longevity of the disc was dramatically confirmed by the detection of O-rich crystalline *silicates* in the mid-infrared spectrum (Waters et al. 1998). The likeliest scenario for O-rich material in this C-rich environment is that silicates in the long-lived circumbinary disk antedate a recent C-rich phase of HD 44179 that expelled the C-rich nebula. Infrared gas-phase $CO_2$ absorptions are also detected towards HD 44179 (Waters et al. 1998). There is even evidence for the fromation of macro structures (Jura & Turner, 1998)

The C-rich extended nebula is best known and often cited to as 'standard' for the well-known but still unidentified infrared emission (UIR) bands at 3.3, 6.2, 7.7, 8.6 and 11.3 $\mu$m (Russell et al., 1978), commonly attributed to Polycyclic Aromatic Hydrocarbons (PAHs: Allamandola et al., 1985).

Although the gross structure of the RR is well-established, the carriers of the extraordinary, presumably molecular, optical

*Send offprint requests to*: Hans Van Winckel,e-mail: Hans.VanWinckel@ster.kuleuven.ac.be.
* Based on observations collected at the European Southern Observatory in Chile (60.C-0473)
** Scientific researcher of the Fund for Scientific Research, Flanders

spectral features in the nebula are still completely unknown. The multi-wavelength panoply of emission lines and bands includes: UV Cameron bands of CO (phosphorescent high-excitation levels likely to arise from the dissociation of $CO_2$), and OH bands (Sitko 1983; Reese & Sitko 1996; Glinski et al. 1996; Glinski & Nuth 1997); Ca H and K; optical $CH^+$ lines (Balm & Jura 1992); the broad extended red emission (ERE), often attributed to photoluminescence of hydrogenated carbon particles (e.g. Witt & Boroson 1990), but also to crystalline nanoparticles of pure silicon (Witt et al., 1998; Ledoux et al., 1998). The latter satisfy the high photoluminescence efficiency needed to explain the observed ERE. Since the ERE is also observed in the interstellar medium, the intrinsic photon conversion efficiency of the photoluminescence must be near 100% (Zubko et al., 1999).

Superposed on the ERE in the RR is a unique and amazingly complex spectrum of intermingled sharp and broad bands (Schmidt et al., 1980; Scarott et al. 1992; Sarre et al. 1995) between roughly 5600 and 7500 Å. The peak wavelengths and widths of some of the features near 5799, 5853 and 6616 Å, were found to decrease with increasing offset from the central HD 44179 and in the limit of observability these were shown to converge toward the narrow diffuse interstellar bands (DIB's) at $\lambda 5797.11$, $\lambda 5849.78$ and $\lambda 6613.72$ Å (Sarre et al., 1995). The profile of these three features, with a steep blue edge and a red degraded tail, is typical of electronic transitions in gas-phase molecules and consists of a rotational branch head, with the redward extensions of unresolved molecular rotational branches that are wider close to the central illuminating source. The observed rotational contours are typical of electronic transitions for which one of the rotational constants is smaller in the excited state than in the ground state. The presence of DIB carriers in the RR was, however, recently questioned by Glinski & Anderson (2001) who found that there remained a wavelength shift far out in the nebula between the DIB profiles and the ERE subfeature profiles.

Since the substructures on the ERE are much richer than the features discussed in detail in the literature, and the identification of the carrier(s) of all the bands would give enormous insight into circumstellar chemistry in C-rich nebula (and hence the ISM), we have studied the ERE and its substructure through long slit spectra with unprecendented S/N. In this article we give a detailed description of the ERE based on high-S/N, medium-resolution, optical ground-based spectra in the nebula. In subsequent papers we will discuss in detail the nebular structure and chemistry based on high-spatial resolution images and long-slit spectra based on HST observations with WFPC2 and STIS, respectively.

After a short description of the observations and reduction methods, we discuss the details of the ERE substructure in Sect. 3. We focus on the DIB analogy in Sects. 4 and 5 and present the spatial changes of the profiles of the subfeatures in Sect. 6. Our main conclusions are summarised in the final section.

## 2. Observations

The 3.5m NTT telescope of ESO at La Silla was used to obtain long-slit spectra of the ERE in several different nebular positions. We used gratings number 6 and 7 with a dispersion of 0.32Å/pix and 0.64Å/pix respectively. The $1''$ slit projected on the sky gave a resolution at 6000Å of $\lambda/\delta\lambda \sim 5500$ and 2600 respectively. The spatial scale is $0.268''$/pix on the sky. For every spectral setting we obtained short integrations on the central object. After a short exposure of the bright central region with slit-position A (P.A.=45°: see Fig. 1), the bright central object was placed just outside the slit to prevent CCD saturation, so we sampled only the north-eastern rim. The integration was split into 3 sub-integrations for the post-integration cosmic ray removal. With the slit in position D (P.A.=105°), we obtained 2 exposures of 1 hour each. The standard reduction included cosmic ray removal, bias subtraction, flatfielding, wavelength calibration and flux response calibration. The sky was subtracted by extrapolating the sky-spectrum from positions outside the nebula. The standard stars used for the response determination were LT3218 and HR3454.

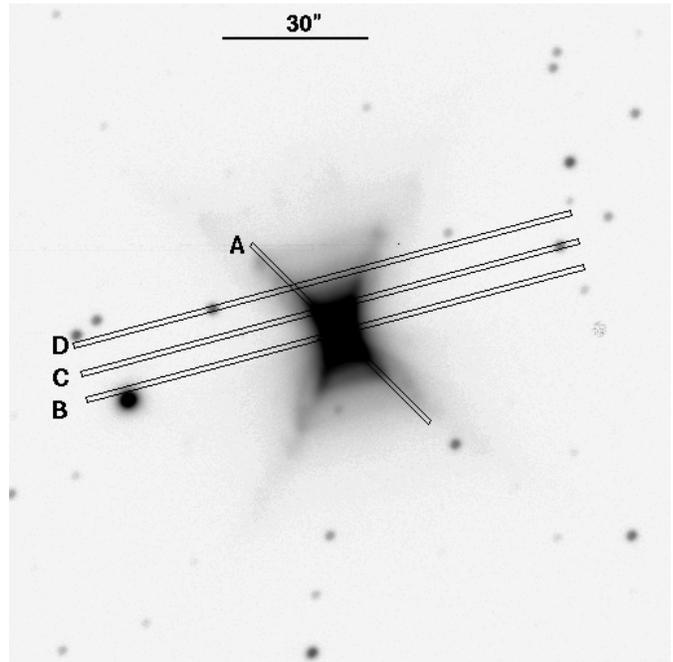

**Fig. 1.** Slit positions of the long-slit observations of the Red Rectangle nebula. The coronographic image is the 20-min exposure through the filter sampling $H_\alpha$. The slit-width was $1''$ and the scale of the CCD is $0.268''$/pix. North is up, East to the left.

**Table 1.** Log of the long-slit observations. The slit positions are given in Fig. 1. The ESO grating number is given with the resolving power obtained and the central wavelength.

| Long-slit spectra NTT | | | | |
|---|---|---|---|---|
| Nr. | R at 6000Å | $\lambda_{cent.}$ Å | range $\Delta\lambda$ | position see Fig 1 | Int. time s |
| 6 | 5500 | 5870 | 640 | A | 180 |
| 6 | 5500 | 5870 | 640 | B | 180 |
| 7 | 2600 | 6200 | 1300 | B | 90 |
| 6 | 5500 | 5870 | 640 | A | 7200 (3 exp.) |
| 6 | 5500 | 5870 | 640 | C ($6''$ offset) | 2700 |
| 7 | 2600 | 6200 | 1300 | D ($11''$ offset) | 7200 (2 exp.) |

**Table 2.** ERE continuum points. We used a smoothed spline function to determine the ERE continuum assuming that the ERE continuum is reached in these windows.

| $\lambda$(Å) | $\lambda$(Å) |
|---|---|
| 5555—5565 | 6130—6140 |
| 5690—5710 | 6145—6153 |
| 5780—5785 | 6298—6542 |
| 5980—5985 | 6607—6611 |
| 6070—6080 | 6750—6860 |

## 3. The ERE spectrum and its substructures

In Fig. 2 and Fig. 3 we show the complete long-slit spectra at $6''$ and $11''$ from the central binary obtained by averaging all the spatial rows sampling the entire nebula. The nebula is $9''$ (34 pixels) wide at $6''$ from the centre, while it is $22''$ wide (82 pixels) at $11''$. The rim of the nebula (along position A in Fig. 1) is brighter in the ERE by $\sim 30\%$ and the subfeatures scale by the same amount. The S/N of the two spectra is about 170.

The ERE is clearly visible as a broad emission hump upon which is superimposed a very complex succession of narrower emission features with a variety of line profiles. The high S/N ratios of the averaged spectra reveal that, besides the already documented emission features, many more but much weaker features are also present for which we give a detailed inventory.

In Table 3 we list the wavelengths of all the narrower ERE subfeatures. All features are clearly resolved except the sodium D-lines. Wavelengths are corrected for the heliocentric systemic velocity of $+18.9$ km s$^{-1}$ as determined by the CO microwave emission (Jura et al. 1995). CO is a much better tracer of the systemic radial velocity since the velocity determined from the optical spectrum of the central object is governed by the viewing angle of the scattering clouds (Waelkens et al., 1996).

The profiles of the bands are complex and we distinguish three types of line profile: 1) the red-degraded (RD) profiles with a steep blue edge and a less sloped red tail (e.g. $\lambda 5799$ Å) (this is a signature of an electron excitation of a free molecule with the extension of unresolved rotational branches towards the red); 2) the often broad, symmetric (S) ones (e.g.: $\lambda 5826.5$ Å); and 3) the irregular (I), where no clear profile can be determined but a jump in the ERE continuum is observed. In Table 3 we list the blue edge (BE, zero intensity level), the peak wavelength (P) and the red edge (RE) of every feature together with the profile class (RD, S, I). For the symmetric profiles we list the FWHM of the Gaussian fit. For the RD profiles, the BE and RE are not always easy to trace, therefore we list the wavelength of the begining and ending of the steep gradient in the profile.

Since the intrinsic profile of the ERE itself is not known, the line profiles of the superimposed emission features are not always clear. This is illustrated in Fig. 3 where a very smooth, interactively determined, continuum is overplotted. We visually determined continuum windows (Table 2) and used a cubic smoothed spline to estimate the ERE continuum. It is clear that, e.g., the $\lambda 6378.6$ Å and $\lambda 6398.8$ Å narrow features are superimposed on a broad emission bump, the strength of which is very dependent on the placement of the underlying continuum. Since the ERE in other sources is found to be rather featureless and smooth (e.g. Witt et al., 1998), we adopt this method, always using the same continuum windows. Assuming this smooth continuum is indeed a good representation of the ERE itself, then from Fig. 3 we see that the emission plateaus are certainly present on top of the ERE.

## 4. DIBs in the ERE of the Red Rectangle?

Fossey (1991) and Sarre (1991) pointed out that one of the proposed families of DIBs appears to correspond in wavelength with a prominent emission band observed in the Red Rectangle. Subsequent more detailed observations (Scarott et al., 1992) revealed that the profiles of the $\lambda 5797$, $\lambda 5850$ and $\lambda 6614$ Å emission features change with distance from the central object. While the blue edge of the feature remains at the same wavelength, the red wing becomes narrower and, at the limit of detectability, the features were found to converge to the DIB wavelengths with similar FWHMs (Scarott et al., 1992). Because the profile and wavelengths of the ERE features correspond to particular sets of DIBs, these authors concluded that the ERE bands originate from the same carrier. The $\lambda 5797$ Å - DIB connection led to quantitative estimates on the possible carrier by e.g. Rouen et al. (1997) and Duley (1998).

This remarkable emission complex around $\lambda 5820$ Å is shown in Fig. 4. The red degraded features at $\lambda 5798$ and $\lambda 5853$ Å have profiles showing a double, resolved, component, while the central $\lambda 5826$ Å broad feature is symmetric. We fitted Gaussians to quantify the different profiles (see Table 4) by central wavelength and FWHM.

Since our spectra have such a high S/N ratio, we can investigate other positional correlations between DIB wavelengths and the ERE subfeatures. In Table 3, we list the wavelengths in the restframe of the RR system having corrected for the $+18.9$ km s$^{-1}$ system velocity. To illustrate the positional agreement between DIBs and ERE subfeatures, we plot a synthetic DIB spectrum obtained in the line-of-sight towards BD+63°1964 (Ehrenfreund et al. 1997) over the normalized ERE (see Fig. 4, 5).

As noted by Sarre et al. (1995) the wavelengths of the blue edges of the red degraded features correspond with the DIBs at $\lambda 5797.11$, $\lambda 5849.78$, $\lambda 613.72$ and possibly 5766.25 Å. Thanks to our high S/N spectra, we did discover more matches between narrow DIB features and DIB wavelengths. In Table 5 we list the wavelength of the blue edge at half maximum in the rest-frame of the RR and the associated DIB wavelength. The new "good" matches correspond to the $\lambda 6196.19$, $\lambda 6203.19$, $\lambda 6234.27$, $\lambda 6445.53$, $\lambda 6709.24$ DIBs. The ERE features are small and they are not all obviously red degraded. The match between the $\lambda 6379.27$ Å DIB and the ERE as discussed in Sarre et al. (1995) is less clear since the ERE is redward of the peak of the feature. Moreover the ERE is complex in that region with evidence of a broad underlying emission plateau (see Fig. 5)

From Fig. 4 is it clear that some other strong DIBs (e.g. $\lambda 5780.59$ Å) are certainly not present in the ERE spectrum.

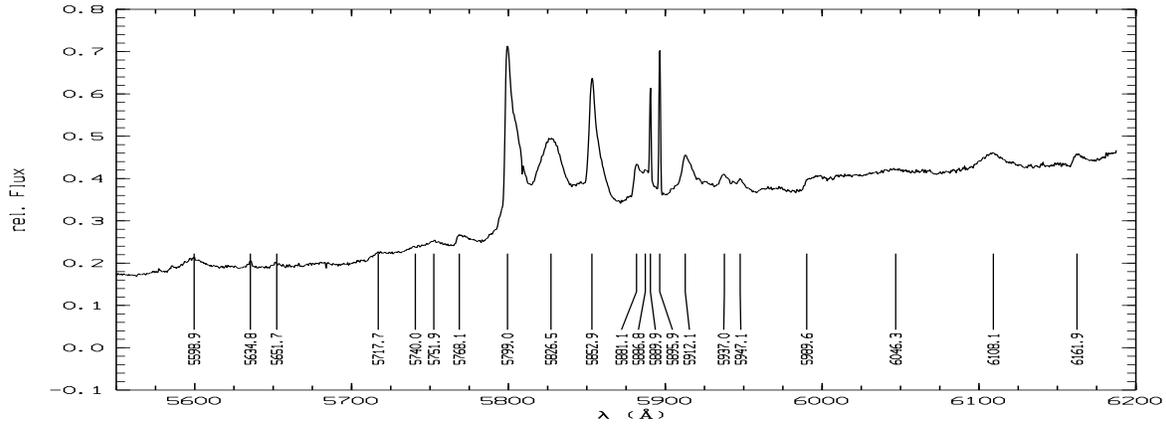

**Fig. 2.** Substructure in the ERE at 6″ from the central object. The spectrum is the mean of all the spectra of the RR nebula obtained along slit-position C. Note the extremely rich and faint substructure on top of the ERE. Wavelengths are corrected for the velocity of the RR for which we adopt a systemic velocity of +18.9 km s$^{-1}$.

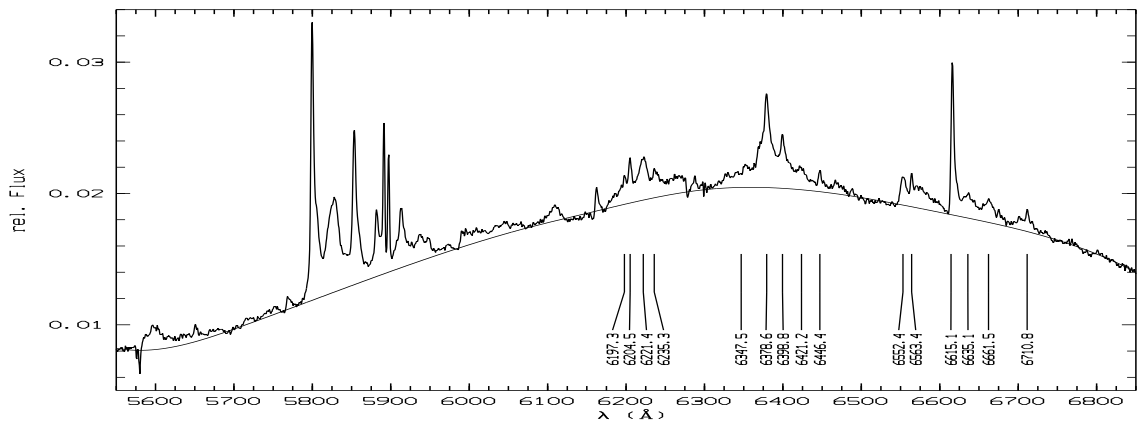

**Fig. 3.** Substructure in the ERE at 11″ from the central object. The spectrum is the mean of all the spectra of the RR nebula obtained along slit-position D. The wavelengths of the features not covered in the higher resolution setting are given, as well as the interactively defined spline through possible continuum points. Wavelengths are corrected for the velocity of the RR system.

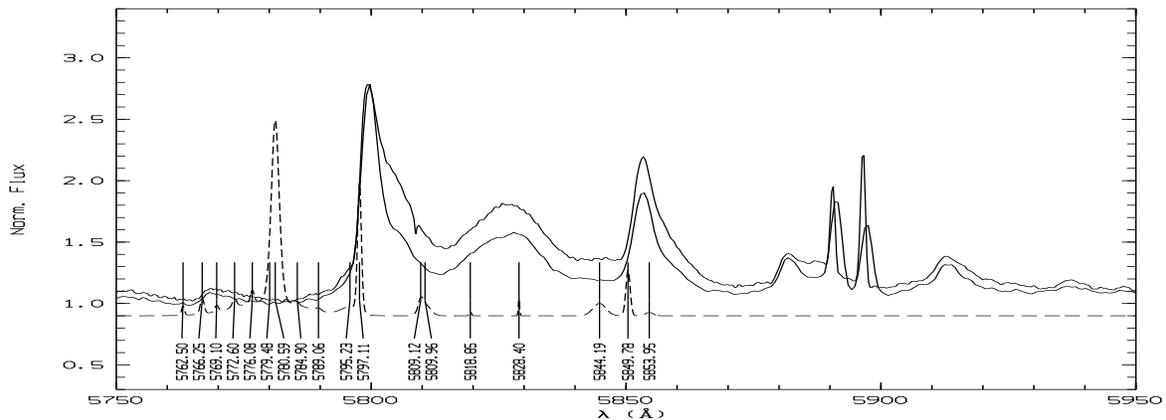

**Fig. 4.** The normalized ERE at an offset of 6″ and 11″ (full lines) from the central object together with the DIB spectrum (broken line) in the direction of BD+63° 1964 as compiled by Ehrenfround et al. (1997). The series of wavelengths is for the DIBs, the DIB spectrum is shifted to correct for the RR system velocity.

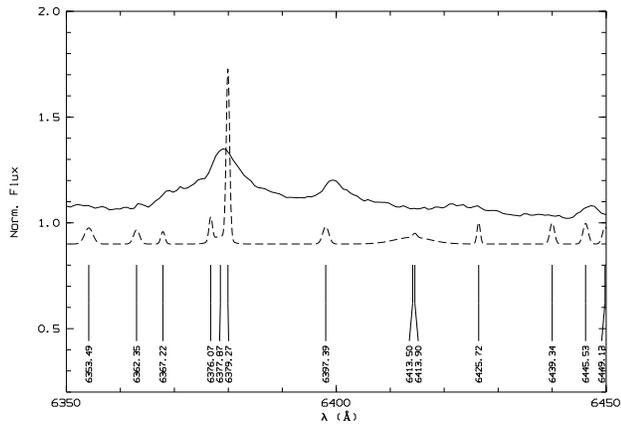

**Fig. 5.** Same as Fig. 4. Only the ERE spectrum at 11″ is shown.

A positional agreement between the blue edge and a DIB certainly does not imply that the DIB carrier is also present in the ERE nebula. For several new candidate DIB/ERE subfeatures, new spectra far from the centre are clearly necessary to confirm these claims. The nebula is, however, faint and such observations need high-quality data with an 8-metre class telescope.

## 5. Spatial information

Schmidt & Witt (1991) found evidence for the displaced maximimum intensity for the ERE compared to the emission subfeatures. They used the integrated line-intensity of the $\lambda 5797$ Å feature and compared this to the monochomatic flux point in the ERE at 6600 Å. This, however, does not address the fact that the ERE itself changes in peak wavelength from close to the object to far out in the nebula (e.g. Witt & Boroson 1990). To check for resulting different morphology of the ERE as compared to the emission features, we integrated the region $\lambda 5787 - 5927$ Å for every spatial pixel on our long-slit spectrum of slit position C. The result is shown in Fig. 6 where we compared the integrated continuum flux (arbitrary units) to the integrated line flux. The latter was multiplied by a factor 3.2. It is clear from Fig. 6 that we do not confirm a displaced maximum of the emission subfeatures compared to the ERE and we find the same conclusion in the slit position D spectrum. In a future paper, we will carefully examine these findings on the basis of high spatial resolution imaging obtained with different filters sampling mainly the ERE and/or the subfeatures.

## 6. Profile changes in the ERE subfeatures

Sarre et al. (1995) described the changes with distance in the RR nebula of the profiles of the ERE subfeatures and noticed the convergence of some features to DIB wavelengths and profiles. This is illustrated in Fig. 7 where normalised slices are shown obtained at position A in the nebula. The vertical lines across the plot correspond to the peak wavelengths of the feature at 2.9″ ($\lambda 5808.10$ Å) and 5.6″ (at $\lambda 5799.3$ Å) from the central object. The profile changes are very clear, referenced to the constant velocity of the NaI D lines. With a constant blue edge and a narrowing red wing, the profile changes of the RD features are typical of unresolved rotational contours of molecules in the changing radiation environment from close to the star, to far out in the nebula. The central wavelength of the symmetric profile does change too but the whole profile shifts to the blue with increasing distance from the central source.

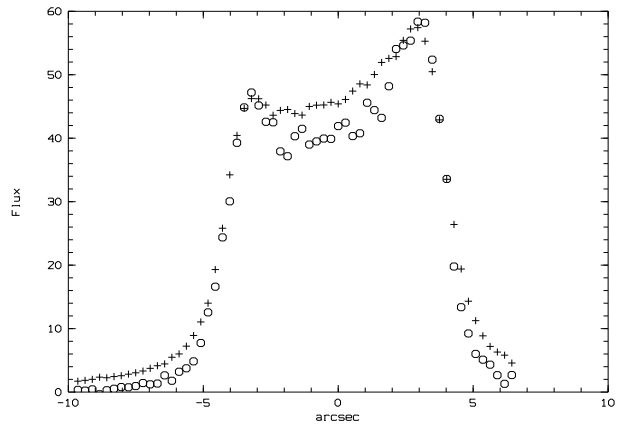

**Fig. 6.** The spatial intensity profiles along the slit position C. The +- signs indicate the continuum flux integrated in the $\lambda 5787 - 5927$ Å window, while the open circles indicate the total line flux in the same wavelength window. For comparison of the profiles, we multiplied the line flux by a factor 3.2

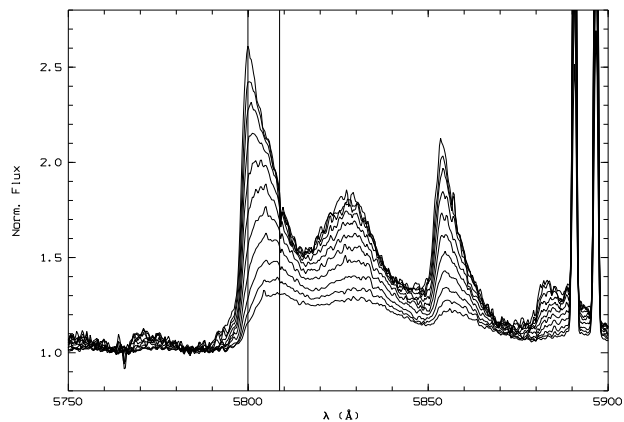

**Fig. 7.** Changing normalised slices through the ERE subfeatures with distance from the central object. From bottom to top are the slices obtained from averaging 2 rows of the long-slit spectra obtained at position A. From bottom to top the profiles are obtained at 2.9″, 3.2″, 3.5″, 3.8″, 4.0″, 4.3″, 4.6″, 4.8″, 5.1″, 5.4″, 5.6″ from the edge of the slit. The vertical lines at $\lambda 5808.1$ and $\lambda 5799.3$ Å correspond respectively to the peak wavelengths of the lower and upper profiles. Wavelengths are measured in the frame of the RR.

We investigated in detail the changes in the peak wavelength of the main $\lambda 5820$ Å complex. The peak wavelengths of the features close to the central object were determined interactively when the feature was strongly non-Gaussian. For the more Gaussian profiles further out in the nebula, the central wavelengths were determined by Gaussian fit-ting. Along the position A slit, we found evidence for the ERE emssion as far as 25″ from the central region.

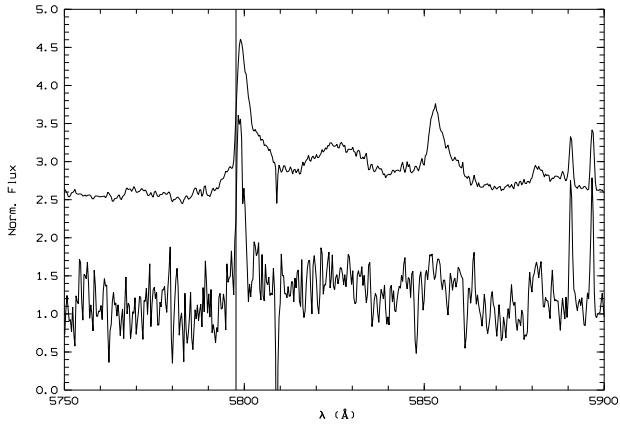

**Fig. 8.** Normalised ERE spectrum at 9.6″ (upper) and 23″ (lower) from the central source measured along the slit (position A). The upper spectrum is offset in flux to see the contrast in wavelength and profile. The full vertical line indicates the expected wavelength of the narrow $\lambda 5797.11$ Å DIB in the frame of the RR system. At 23″ this feature is much stronger than the invisible $\lambda 5849$ Å feature. On the top panel the double component of the resolved $\lambda 5797$ Å feature is clear.

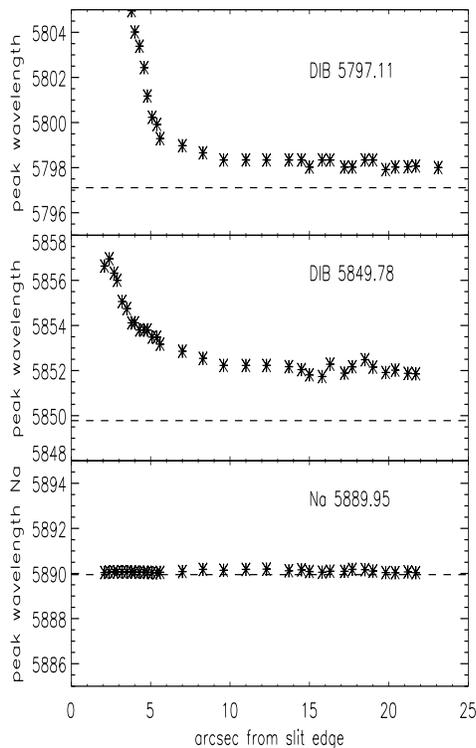

**Fig. 9.** The top 2 panels show the wavelength shifts of the peak of the emission features against the distance from the central source. The wavelength drop during the first ∼10″ is very clear but then the wavelength levels off and even at 24″ along the slit, the peak wavelength does not converge to the DIB reference wavelength.

We plot the peak wavelengths in Fig. 9 of both red-degraded features (the $\lambda 5797$ and 5849 Å) together with the wavelengths determined for the Na D line at $\lambda 5889.95$Å. In the first 5-10″ the profile changes dramatically, with a peak wavelength shift of at least 6 Å. Then the wavelength levels off to a constant value which is slightly redward of the DIB wavelength. For the $\lambda 5798$ Å feature, the central wavelength determination at 24″ from the edge of the slit is still 1 Å different from the DIB-wavelength while the $\lambda 5849$ Å feature converges to a wavelength 2 Å redward of the DIB wavelength. From the consistent Na line meaurements it is clear that this shift is certainly not due to wavelength calibration errors. The same shift was independently detected in lower S/N but higher resolution spectra discussed by Glinski & Anderson (2002).

There is growing observational evidence for the molecular origin of the DIBs (e.g. see Furla & Krełowski 2000) and the spatial profile changes of the red degraded bands in the RR also show a molecular signature at very similar wavelength positions.

## 7. Discussion

The main conclusions of the ERE long-slit spectra can be summarised as follows : 1) The ERE substructure is far richer than the set of strong features described in the literature. A detailed inventory of the features is offered in Table 3. Assuming a smooth continuum, we have evidence for strong plateaus of emission. 2) Additional features have wavelength positions in which the blue edge corresponds with a DIB wavelength. Specifically, we propose 5 new subfeatures with a possible DIB connection besides the already documented ones. 3) Detailed spatially-resolved spectra show, however, that the main emission features at $\lambda 5797$ and 5849 Å do not converge to DIB wavelengths, even 24″ from the central source, but remain 1 respectively 2 Å redwards of the DIB wavelength. The complex ERE spectral subfeatures show that the chemistry in the RR is indeed very rich. A focused high (spatial) resolution inventory of all the features, and their changing profiles from close to the central object to far out in the faint nebula, is certainly warranted to impose observational constraints from rotational contour studies.

In subsequent papers we will discuss in detail the spectro-spatial behaviour of the bands in great depth, together with a detailed study of the structure of the RR, from close to the central object to far out in the nebula.

*Acknowledgements.* The authors thank Jan Cami for the digital DIB information and the discussions concerning DIBs and the anonymous referee for the interesting remarks. HVW acknowledges financial support from the Fund for Scientific Research of Flanders. MC thanks NASA for its support of his participation in these studies under grant HST-GO-07297.01-A with Berkeley.

**Table 3.** The wavelengths of the ERE-subfeatures detected at 6″ and 11″ from the central binary. Wavelengths are corrected for the RR systemic velocity. For the symmetric profiles (S) we list the peak wavelength (P) and the FWHM of the Gaussian fit. For the red degraded profiles (RD) we list the blue edge (BE) at zero intensity, the peak wavelength (P) and the red edge (RE). The irregular features (I) have no clear profile.

| $\lambda$ 6″ | $\lambda$ 11″ | Position | FWHM Å | Profile |
|---|---|---|---|---|
| 5598.9 | 5598.3 | P | 17,14 | S |
| 5634.8 | | P | 2 | S |
| 5651.7 | 5650.6 | P | 5,4 | S |
| 5717.7 | | P | 10 | S |
| 5740.0 | | P | 10 | S |
| 5751.9 | | P | 10 | S |
| 5764.6 | 5763.7 | BE | | I |
| 5768.1 | 5767.7 | P | | |
| 5781.3 | 5777.5 | RE | | |
| 5795.8 | 5794.9 | BE | | RD |
| 5799.0 | 5799.2 | P | | |
| 5812.5 | 5811.0 | RE | | |
| 5826.5 | 5826.4 | P | 13 | S |
| 5848.4 | 5848.3 | BE | | RD |
| 5852.9 | 5852.9 | P | | |
| 5865.7 | 5865.4 | RE | | |
| 5881.1 | 5881.4 | S | 4,4 | S |
| 5886.8 | | S | 9 | S |
| 5889.92 | 5890.58 | Na I | | S |
| 5895.86 | 5896.62 | Na I | | S |
| | 5907.4 | BE | | RD |
| 5912.1 | 5913.4 | P | | |
| 5919.3 | | RE | | |
| 5937.0 | 5937.2 | P | 7 | S |
| 5947.1 | 5947.0 | P | 4 | S |
| 5985.8 | | BE | | I |
| 5989.6 | | P | | |
| 6046.3 | | P | 24 | S |
| 6108.1 | 6108.3 | P | 17,16 | S |
| 6157.2 | 6157.3 | BE | | RD |
| 6161.9 | 6161.9 | P | | |
| 6170.2 | 6169.9 | RE | | |
| | 6195.3 | BE | | RD |
| | 6197.30 | P | | |
| | 6199.9 | RE | | |
| | 6200.6 | BE | | RD |
| | 6204.5 | P | | |
| | 6208.4 | RE | | |
| | 6221.4 | S | 10 | S |
| | 6233.3 | BE | | RD |
| | 6235.3 | P | | |
| | 6241.2 | RE | | |
| | 6347.5 | | | I |
| | 6378.6 | P | 5 | S |
| | 6398.8 | P | 4 | S |
| | 6421.2 | P | 18 | S |
| | 6446.4 | P | 3 | S |
| | 6552.4 | P | 7 | S |
| | 6563.4 | P | | * |
| | 6611.1 | BE | | RD |
| | 6615.1 | P | | |
| | 6624.9 | RE | | |
| | 6635.1 | P | 14 | S |
| | 6661.5 | P | 11 | S |
| | 6710.8 | P | 3 | S |

* : skyline confusion

**Table 4.** Gaussian decomposition of the emission complex around $\lambda 5820$ Å. The RD features at $\lambda 5797$ and $\lambda 5850$ are described with a double Gaussian fit with a narrow component and a red-displaced broad component.

| component | $\lambda$ (Å) | FWHM (Å) |
|---|---|---|
| 1 | 5798.8 | 3.1 |
| 2 | 5802.7 | 10.4 |
| 3 | 5825.6 | 23.1 |
| 4 | 5852.8 | 3.5 |
| 5 | 5854.7 | 12.8 |

**Table 5.** DIB-ERE correlation reveales the presence of more candidate DIB features in the complex emission features of the ERE. We list the measured wavelength of the blue edge at half maximum in the Red Rectangle restframe of the spectrum at 11″ together with the associated DIB wavelength.

| ERE | profile | DIB |
|---|---|---|
| 5765.7 | I | 5766.25* |
| 5797.1 | RD | 5797.11* |
| 5850.6 | RD | 5849.78* |
| 6169.3 | RD | 6196.19 |
| 6202.6 | RD | 6203.19 |
| 6234.2 | RD | 6234.27 |
| 6444.5 | S | 6445.53 |
| 6613.1 | RD | 6613.72* |
| 6709.5 | S | 6709.24 |

* : see Sarre et al., 1995